\documentclass[11pt]{article}
\usepackage{graphicx,amsmath}

\newcommand{\cent}{c\hspace{-1.5mm}/}

\newcommand{\bsquare}{\hbox{\rule{6pt}{6pt}}}
\newcommand{\condi}{w_1=w_2^R}
\newcommand{\condin}{w_1 \neq w_2^R}
\newcommand{\condii}{(w_1w_2) =(w_3 w_4)^R}
\newcommand{\condiin}{(w_1w_2) \neq (w_3 w_4)^R}

\newcommand{\frai}{\frac{N_0}{N_1}}
\newcommand{\fraii}{\frac{N_0'}{N_2}}
\newcommand{\sumk}{\sum_{k=1}^{N_1}}

\newcommand{\suml}{\sum_{l=1}^{N_2}}

\newcommand{\qi}{|q_{p_k,e,p_l,f,1}\rangle}
\newcommand{\qii}{|q_{p_k,e,p_l,f,2}\rangle}
\newcommand{\qiii}{|q_{p_k,e,p_l,f,3}\rangle}
\newcommand{\qiiii}{|q_{p_k,e,p_l,f,4}\rangle}
\newcommand{\ei}{0\le e < \frac{p_k}{2}}
\newcommand{\eii}{\frac{p_k}{2} < e < p_k}
\newcommand{\eiii}{0\le e < \frac{p_k}{2}-1}
\newcommand{\eiiii}{\frac{p_k}{2}-1 < e < p_k}
\newcommand{\eeven}{e: \ even}
\newcommand{\eodd}{e: \ odd}
\newcommand{\ffi}{0\le f < \frac{p_l}{2}}
\newcommand{\ffii}{\frac{p_l}{2} < f < p_l}
\newcommand{\ffiii}{0\le f < \frac{p_l}{2}-1}
\newcommand{\ffiiii}{\frac{p_l}{2}-1 < f < p_l}
\newcommand{\feven}{f: \ even}
\newcommand{\fodd}{f: \ odd}
\begin{document}






\begin{center}
{\LARGE Exploiting the Difference in Probability
  Calculation between Quantum and Probabilistic Computations}\\
\end{center}

\begin{center}
Masami Amano, Kazuo Iwama and Rudy Raymond H.P.\\
Graduate School of Informatics, Kyoto University\\
\{masami, iwama, raymond\}@kuis.kyoto-u.ac.jp
\end{center}

\vspace{-5mm}

\begin{abstract}
  The main purpose of this paper is to show that 
we can exploit
the difference ($l_1$-norm and $l_2$-norm) in the probability
calculation between quantum and probabilistic computations to claim
the difference in their space efficiencies. It is shown that
there is a finite language $L$ 
which contains sentences of length up to $O(n^{c+1})$ such that:
($i$) There is a one-way quantum finite automaton (qfa) of
$O(n^{c+4})$ states which recognizes $L$. ($ii$) However, if
we try to simulate this
qfa by a probabilistic finite automaton (pfa) \textit{using the
  same algorithm}, then it needs $\Omega(n^{2c+4})$ states. It
should be noted that we do not prove real lower bounds for pfa's
but show that if pfa's and qfa's use exactly the same algorithm, 
then qfa's need much less states.
\end{abstract}

\vspace{-5mm}

\section{Introduction}
\vspace{-2mm}

 It is a fundamental rule of quantum computation that if a state $q$
has an amplitude of $\sigma$, then $q$ will be observed not with
probability $\|\sigma\|$ but with probability
$\|\sigma\|^2$. Therefore, if one can increase the amplitude of $q$
twice, i.e., from $\sigma$ to $2\sigma$, then the corresponding
probability increases four times, i.e., $\|\sigma\|^2$ to
$4\|\sigma\|^2$. In general, if the amplitude increases $k$ times then
the probability increases $k^2$ times. One observation of the Grover
search \cite{Gro96}, which is apparently one of the most celebrated
quantum algorithms, is that it takes advantage of this fact cleverly,
by inventing the (efficient) quantum process whose $k$ iterations
increase the amplitude of a designated state roughly $k$ times. As
described above, this is equivalent to increasing the probability
$k^2$ times. Thus the Grover search is faster quadratically than the
classic randomized search whose $k$ iterations can increase the
probability only $k$ times. 

In this paper, we also exploit this feature, i.e., the difference in
probability calculation between quantum and probabilistic
computations, but from a bit different angle: Suppose that there are
ten pairs of state $(p_1, q_1),\cdots,(p_{10},q_{10})$ where, for each
$1 \le  i \le 10$, either $p_i$ or $q_i$ has the amplitude 
$1/\sqrt{10}$ (we say that $p_i$ is ON if it has the amplitude
and OFF otherwise.). We wish to know how many $p_i$'s are
ON. This can be done by ``gathering'' amplitudes by applying a
Fourier transform from $p_i$'s to $r_i$'s and by observing
$r_{10}$ (see later sections for details). If all ten $p_i$'s
are ON, then the amplitude of
$r_{10}$ after Fourier transform is one and it is observed
with probability one. If, for example, only three $p_i$'s are
ON, then the amplitudes of $r_{10}$ is 3/10 and is observed with 
probability 9/100. In the case of probabilistic computation, we can
also gather the  probability of $p_i$'s ($=$ 1/10 for each) simply by
defining (deterministic) transitions from $p_i$ to $r_{10}$. If all
pairs are ON, then the probability that $r_{10}$ is observed is one
again, but if only three $p_i$'s are ON, its probability is 3/10.
If the latter case (only three $p_i$'s are ON) is associated
with some erroneous situation, this probability, 3/10, is much
larger than 9/100 in the quantum case. In other words quantum
computation can enjoy much smaller error-probability due to 
the difference in the rule of probability calculation.

The question is of course whether we can turn this feature into
some concrete result or how we can translate this difference in
probability into some difference in efficiency like time and
space. In this paper we give an affirmative answer to this
question by using 
quantum finite automata; we prove that there is a finite language $L$ 
which contains sentences of length up to $O(n^{c+1})$ such that: 
($i$) There is a one-way quantum finite automaton (qfa) of
$O(n^{c+4})$ states which recognizes $L$. ($ii$) However, if
we try to simulate this 
qfa by a probabilistic finite automaton (pfa) \textit{using the
  same algorithm}, then it needs $\Omega(n^{2c+4})$ states. It
should be noted that we do not prove real lower bounds for pfa's 
but show that if pfa's and qfa's use exactly the same algorithm
(the only difference is the way of gathering amplitudes 
mentioned above), then qfa's need much less states. 

Quantum finite automata have been popular in the
literature since its simplicity is nice to understand merits and
demerits of quantum
computation\cite{AF98,AG00,AI99,ANTV99,KW97,Nay99}. Ambainis and
Freivalds \cite{AF98} proved an exponential difference in the size of
qfa's and pfa's for a one-letter language. Their technique highly
depends on the \textit{rotation of complex amplitudes}, which is
exceptionally powerful for a certain situation. Nayak\cite{Nay99} gave
a negative side of qfa's by showing $L_n = \{wa|\ w \in \{a,b\}^{*}\ 
and\ |w| \le n\}$ needs exponentially more states for qfa's than for
dfa's. $L_{\infty}$ is a regular set but is not recognizable by qfa's
as shown by Kondacs and Watrous\cite{KW97}. \cite{KW97} also
introduced 2-way qfa's which can recognize non-regular languages. To
our best knowledge, the notion of gathering amplitudes using Fourier
transform appeared in this paper for the first time and played an
important role in \cite{AI99}, too.   

\vspace{-5mm}
\section{Problem EQ}
\vspace{-2mm}

Suppose that Alice and Bob have $n$-bit numbers $x$ and $y$ and
they wish to know whether or not $x=y$. This problem, called EQ, 
is one of the most famous problems for which its randomized
communication complexity ($=\Theta(\log n)$) is significantly
smaller than its deterministic counterpart
($=n+1$) \cite{KN97}. In this paper, we need a little bit more
accurate 
argument on the value of randomized (and one-way) communication
complexity: Consider the following protocol $M_{EQ}$: ($i$)
Alice selects a single prime $p$ among the smallest $N$
primes. ($ii$) Then she divides $x$ by $p$ and sends Bob $p$ and 
the residue $a$. ($iii$) Bob also divides his number $y$ by $p$
and compares his residue with $a$. They accept $(x,y)$ iff those 
residues coincide.

It is obvious that if $x=y$ then protocol $M_{EQ}$ accepts
$(x,y)$ with probability one. Let $E(N)$ be the maximum (error)
probability that $M_{EQ}$ accepts $(x,y)$ even if $x \neq y$. To 
compute $E(N)$, we need the following lemma: (In this paper,
$\log n$ always means $\log_2 n$ and $\lceil f(n) \rceil$ for a
function $f(n)$ is simply written as $f(n)$.)

\textbf{Lemma 1}. Suppose that $x \neq y$ and let $S(x,y)$ be a set
of primes such that $x=y$ mod $p$ for all $p$ in $S(x,y)$. Also, let 
$s$ be the maximum size of such a set $S(x,y)$ for a pair of $n$-bit
integers $x$ and $y$. Then $s=\Theta(n/\log n)$.

\textbf{Proof}. Let $p_i$ be the $i$-th largest prime and
$\pi(n)$ be the number of different primes $\le n$. Then the
prime number theorem says that $\lim_{n \to
  \infty}\frac{\pi(n)}{n/\log_e n}=1$, which means that $p_{n/\log 
  n}= \Theta(n)$. Consequently, there must be a constant $c$ s.t. $p_1 \cdot p_2 \cdots p_{n/\log n} \cdots
p_{cn/\log n} > p_{n/\log n}\cdot p_{n/\log n+1}\cdots p_{cn/\log n}> 2^n$
since $n^{n/\log n} = 2^n$. Thus an 
$n$-bit integer $z$ has at most $cn/\log n$ different prime
factors. Note that $x=y$ mod $a$ iff $|x-y|=0$ mod
$a$. Hence, $s \le cn/\log n$. Also it turns out by the prime number
theorem that there is an $n$-bit integer $z$ such that it has
$c'n/\log n$ different prime factors for some constant $c'$,
which proves that $s \ge c'n/\log n$. \hfill \bsquare

In this paper, $N_0$ denotes this number $s$ which is
$\Theta(n/\log n)$. Then

\textbf{Lemma 2}. $E(N) = N_0/N$.

For example, if we use $N=n^2/\log n$ different primes in $M_{EQ}$,
its error-rate is $1/n$.

\vspace{-5mm}
\section{Our Languages and qfa's}
\vspace{-2mm}

A one-way qfa is the following model: ($i$) Its input head
always moves one position to the right each step. ($ii$) Global
state transitions must be unitary. ($iii$) Its states are
partitioned into \textit{accepting}, \textit{rejecting} and
\textit{non-halting} states. ($iv$) 
Observation is carried out every step, and if acceptance or
rejection is observed, then the computation halts. Otherwise,
computation continues after proportionally distributing the amplitudes
of accepting and rejecting states to non-halting states. We omit
the details, see for example \cite{KW97}. In this paper, we
consider the following three finite languages.

$L_0(n)=\{w\sharp w^R \mid w \in \{0,1\}^n\},$

$L_1(n)=\{w_1 \sharp w_2 \sharp \sharp w_3 \sharp w_4 \sharp \mid
w_1,w_2,w_3,w_4 \in \{0,1\}^n,(\condi)\vee(\condii)\},$

$L_2(n,k)=\{w_{11}\sharp w_{12} \sharp \sharp
w_{13} \sharp w_{14} 
\sharp \sharp \sharp \cdots \sharp \sharp \sharp w_{i1} \sharp
w_{i2} \sharp \sharp w_{i3} \sharp w_{i4} \sharp \sharp \sharp
\cdots \sharp \sharp \sharp w_{k1} \sharp w_{k2} \sharp \sharp
w_{k3} \sharp w_{k4} \sharp \mid$\\
\hspace{2.5cm}$w_{i1}$, $w_{i2}$, $w_{i3}$, $w_{i4}$
$\in\{0,1\}^n$,\ $1\le i \le k$\\
\hspace{2.5cm}and $1\le \exists j \le k$
s.t. ($w_{j1}=w_{j2}^R$) $\wedge$ (for all $1 \le i \le 
j-1$, $(w_{i1}w_{i2})=(w_{i3}w_{i4})^R$)\}.

In the next section, we first construct a qfa $M_0^Q$, which
accepts each string $x \in L_0$ with probability 1 and each string $y
\notin 
L_0$ with probability at most $\frac{1}{n}$. $M_0^Q$ simulates the 
protocol $M_{EQ}$ in the following way (see Fig 1). Given an 
input string $\cent w_1 \sharp w_2 \$ $ ($\cent$ is the leftmost 
and \$ is the rightmost symbols), $M_0^Q$ first splits into $N$
different states $q_{p_1},\cdots,q_{p_i},\cdots,q_{p_N}$ with
equal amplitudes by reading $\cent$. Then from $q_{p_i}$,
submachine $M_{1i}$ starts the task for dividing integer $w_1$ by the $i$-th prime
$p_i$. This computation ends up in some state of $M_{1i}$ which
corresponds to the residue of the
division. This residue is shifted to the next
submachine $M_{2i}$, and then $M_{2i}$ carries out a completely
opposite operation while reading $w_2$. If (and only if) two
residues
are the same, $M_{2i}$ ends up in some specific state
$q_i^0$. $M_0^Q$ then applies a Fourier transform from
$q_i^0$ to $s_i$ for $1 \le i \le N$. $M_0^Q$ thus simulates
$M_{EQ}$ by setting $s_N$ as its only accepting state.

\begin{figure}[htbp]
 \begin{center}
  \scalebox{0.4}{\includegraphics{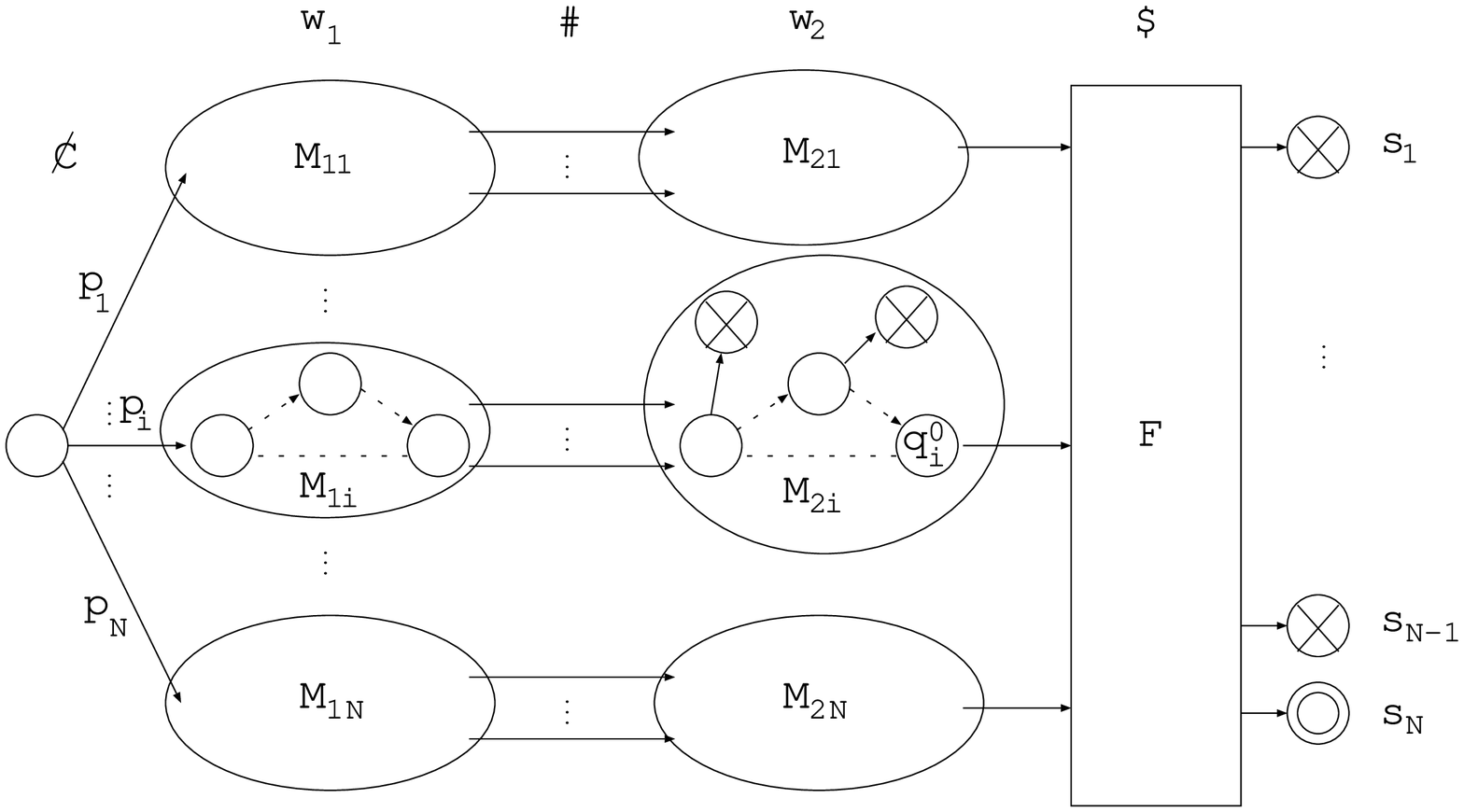}}
  
  \textbf{Fig 1. qfa $M_0^Q$}

\vspace{-0.8cm}
 \end{center}
\end{figure}

For the probabilistic counterpart, pfa $M_0^P$, we can use exactly the
same state transition, except for deterministic transitions from $q_i^0$ to $s_N$. As 
mentioned before we can achieve a quadratic difference in the
probability of error, like $(1/n)^2$ for $M_0^Q$ v.s. $(1/n)$ for
$M_0^P$. It would be nice if this quadratic difference of error can
be traded directly to a quadratic difference in the 
necessary number of primes or to a quadratic difference
in the size of automata. Unfortunately that is not possible: The main reason
is that we do not need such a small (like $1/n$ or $1/n^2$) error-rate
but something like 1/3 is enough by definition. Then the quadratic
difference in the error is reduced to a difference between, say, 1/3
and 1/9, which means only a difference of the constant
factor in the necessary number of primes or the necessary number of
states.

There is a standard technique to overcome this difficulty,
namely, the use of iteration. Consider the following string:
$$w_{11}\sharp w_{12} \sharp \sharp w_{21} \sharp w_{22} \sharp
\sharp \cdots \sharp w_{n1} \sharp w_{n2}$$
where the accepting condition is that for some $1 \le j \le n$,
$w_{j1}=w_{j2}^R$. When all pairs $(w_{j1},w_{j2})$ do not
satisfy this condition, the (error) probability of accepting
such a string is roughly $O\left(\frac{1}{n}\right)\times n =
O(1)$, which appears desirable for our purpose.

This argument does not seem to cause any difficulty for pfa's but
it does for qfa's for the following reason: After checking
$w_{11}$ and $w_{12}$, the qfa is in a single accepting state if 
the condition is met, which is completely fine. However, if
$w_{11}\neq w_{12}^R$ and the observation is not accepting, then 
there are many small amplitudes distributed to many different
states. Note that we have to continue the calculation for
$w_{21}$ and $w_{22}$ which should be started from a single
state. (It may be possible to start the new computation from
each non-halting state, but that will result in an exponential
blow-up in the number of states, which again implies no clear
separation in the 
size of automata.) One can see easily that we cannot use a
Fourier transform this time to gather the amplitudes since there 
are many different patterns in the distribution of states which
have a small nonzero amplitudes.

This is the reason why the next language $L_1(n)$ plays an important
role. Suppose that $\condin$. Then the resulting distribution of 
amplitudes is quite complicated as mentioned above. However,
no matter how it is complicated, we can completely ``reverse'' the
previous 
computation for $w_1 \sharp w_2$ by reading $w_3 \sharp w_4$ if
$(w_1w_2)=(w_3w_4)^R$. This reverse computation should end up in
a single state of amplitude one (actually it is a little less
than one) since the original computation for $\condin$ starts
from the (single) initial state. Now one can use the iteration scheme,
which naturally leads us to the third language $L_2(n,k)$.

\vspace{-5mm}
\section{Main Results}
\vspace{-2mm}

As mentioned in the previous section, we construct
our qfa's and corresponding pfa's for $L_0(n)$, $L_1(n)$ and
$L_2(n,n^c)$ in a step-by-step fashion. Recall that $N$ is the number
of primes used in 
protocol $M_{EQ}$ and $N_0=\Theta(n/\log n)$.

\textbf{Lemma 3}. There exists a qfa $M_0^Q$ which accepts
strings in $L_0$ with probability one and strings not in $L_0$
with probability at most $\left(\frac{N_0}{N}\right)^2$. The
number of states in $M_0^Q$ is $\Theta(N^2\log N)$.

\textbf{Proof}. $M_0^Q$ has the following states: ($i$) An
initial state $q_0$, ($ii$) $q_{p_k,j_k,1}$ (in submachine
$M_{1i}$ of Fig 1), ($iii$) $q_{p_k,j_k,2}$ (in $M_{2i}$ of Fig
1), ($iv$) $q_{p_k,j_k,rej}$ (also in $M_{2i}$ of
Fig 1), ($v$) $s_l$, where $1 \le k \le N$, $0 \le j_k \le
p_k-1$ and $1 \le l \le N$. $p_k$ denotes the $k$-th largest prime
$>3$ (two is excluded for the reason mentioned later). $s_N$ is
the only accepting state, $q_{p_k,j_k,rej}$ and $s_l$ ($1\le l
\le N-1$) are rejecting states and all the others are non-halting
states. We give a complete state transition diagram of $M_0^Q$
in Table 1, where $V_\sigma|Q\rangle=\alpha_1|Q_1\rangle +
\cdots + \alpha_i|Q_i\rangle+\cdots+\alpha_m |Q_m\rangle$ means
that if $M_0^Q$ reads symbol $\sigma$ in state $Q$, it moves to
each state $Q_i$ with amplitude $\alpha_i$ ($|\alpha_1
|^2+\cdots +|\alpha_m |^2=1$).

When reading $\cent$ of the input string $\cent w_1 \sharp w_2
\$$, $M_0^Q$ splits into $N$ submachines (denoted by $M_{1i}$ in
Fig 1) with equal amplitudes (see transition (1) of Table
1). The $k$-th submachine $M_{1k}$ computes the residue when
dividing $w_1$ by $p_k$ (by using transition $(2-a)$ to $(2-d)$ in
Table 1). This division can be done simply by simulating the
usual division procedure as shown in Fig 2 ($a$) and ($b$) for
$w_1=110001$ and $p_2=101$ ($=5$). State $j$ in Fig 2 ($b$)
corresponds to $q_{p_2,j,1}$. The starting state is 0 and by
reading the first symbol 1 it goes to state 1. By reading the
second symbol 1, it goes to state 3 ($=11$). Now reading 0, it
goes to state 1
since $110=1$ mod 101. This continues until reading the last
symbol 1 and $M_0^Q$ ends up in state 4. It should be noted that 
these state transitions are reversible: For example, if the
machine reaches
state 2 ($=10$) from some state $Q$ by reading 0, then $Q$ must be
state 1 since $Q$ cannot be greater than 2. (Reason: If $Q$ is
greater than
2, it means that the quotient will be 1 after reading a new
symbol. Since $M_0^Q$ reads 0 as the new symbol,
the least significant bit of the residue when divided by 5 must
be 1, which excludes state 2 as its next state.) Hence the
quotient must have been
0, and so the previous state must be 1. Note that this argument 
holds because we excluded two, which is the only even prime, from
$p_k$.

Thus, if $w_1$ mod $p_k=j_k$, then $M_0^Q$ is in superposition
$\frac{1}{\sqrt{N}}\sum_{k=1}^{N}|q_{p_k,j_k,1}\rangle$ after
reading $w_1$. Then $M_0^Q$ reads $\sharp$ and this superposition
is ``shifted'' to $\frac{1}{\sqrt{N}}$ $\sum_{k=1}^{N}|q_{p_k,j_k,2}
\rangle$, where $M_0^Q$ checks if $w_2^R$ mod $p_k$ is also
$j_k$ by using transition $(4-a)$ to $(4-d)$ in Table 1. This job
can be done by completely reversing the previous procedure of
dividing $w_1$ by $p_k$. Actually, the state transitions are
obtained by simply reversing the directions of previous state
diagrams. Since previous transitions are reversible, new
transitions are also reversible. Now one can see that the $k$-th 
submachine $M_{2k}$ is in state $q_{p_k,0,2}$ iff the two
residues are the same.

Finally by reading $\$$, Fourier transform is carried out
only from these zero-residue states $q_{p_k,0,2}$ to
$s_l$. From other states $q_{p_k,j,2}$ ($j
\neq 0$) $M_0^Q$ goes to rejecting states $q_{p_k,j,rej}$. If the
residues  
are the same in only $t$ submachines out of the $k$ ones, the
amplitude of $s_N$ is computed as
$$\frac{1}{N}\sum_{|t|}\sum_{l=1}^N\exp\left(\frac{2\pi
    i}{N}kl\right)|s_l\rangle=\frac{t}{N}|s_{N}\rangle +
    \frac{1}{N}\sum_{|t|}\sum_{l=1}^{N-1}\exp\left(\frac{2\pi 
    i}{N}kl\right)|s_l\rangle,$$
namely that is equal to $t/N$. Thus the probability of acceptance is
$\left(\frac{t}{N}\right) ^2$. If the input string is in $L_0$,
then this probability becomes 1. Otherwise, it is at most
$\left(N_0/N\right)^2$ by Lemma 2. The number of states in
$M_0^Q$ is given as
$$
1 + 2\sum_{k=1}^{N} p_k + \sum_{k=1}^{N} (p_k-1) + N 
= 1 + 3 \sum_{k=1}^{N} p_k 
\le 1 + 3 \cdot N \cdot p_{N} 
= O(N^2 \log N),
$$
which completes the proof. \hfill \bsquare

  \begin{center}
$$
\begin{array}{llll}
(1) & V_{\cent}|q_0\rangle = \frac{1}{\sqrt{N}}
\sum_{k=1}^{N}|q_{p_k,0,1}\rangle, & (4-a) & V_0|q_{p_k,j,2}\rangle =
|q_{p_k,\frac{j}{2},2}\rangle \quad (j: \ even), \\

(2-a) & V_0|q_{p_k,j,1}\rangle = |q_{p_k,2j,1}\rangle \quad (0 \le j <
\frac{p_k}{2}), & (4-b) & V_0|q_{p_k,j,2}\rangle =
|q_{p_k,\frac{j+p_k}{2},2}\rangle \quad (j: \ odd), \\

(2-b) & V_0|q_{p_k,j,1}\rangle = |q_{p_k,2j-p_k,1}\rangle \quad
(\frac{p_k}{2} < j < p_k), & (4-c) & V_1|q_{p_k,j,2}\rangle =
|q_{p_k,\frac{j-1+p_k}{2},2}\rangle \quad (j: \ even), \\ 

(2-c) & V_1|q_{p_k,j,1}\rangle = |q_{p_k,2j+1,1}\rangle \quad (0 \le j
< \frac{p_k}{2}-1), & (4-d) & V_1|q_{p_k,j,2}\rangle =
|q_{p_k,\frac{j-1}{2},2}\rangle \quad (j: \ odd), \\ 

(2-d) & V_1|q_{p_k,j,1}\rangle = |q_{p_k,2j+1-p_k,1}\rangle \quad
(\frac{p_k}{2}-1 < j < p_k), & (5-a) & V_\$|q_{p_k,0,2}\rangle=\frac{1}{\sqrt{N}}
\sum_{l=1}^{N}\exp\left(\frac{2\pi i}{N}kl\right)|s_l\rangle, \\ 

(3) & V_{\sharp}|q_{p_k,j,1}\rangle = |q_{p_k,j,2}\rangle, & (5-b) &
V_\$|q_{p_k,j,2}\rangle=|q_{p_k,j,rej}\rangle \quad (1 \le j < 
p_k). 
\end{array}
$$
\vspace{-0.5cm}
    \textbf{Table 1.} State transition diagram of $M_0^Q$
  \end{center}

\begin{figure}[htbp]
  \begin{center}
  \scalebox{0.45}{\includegraphics{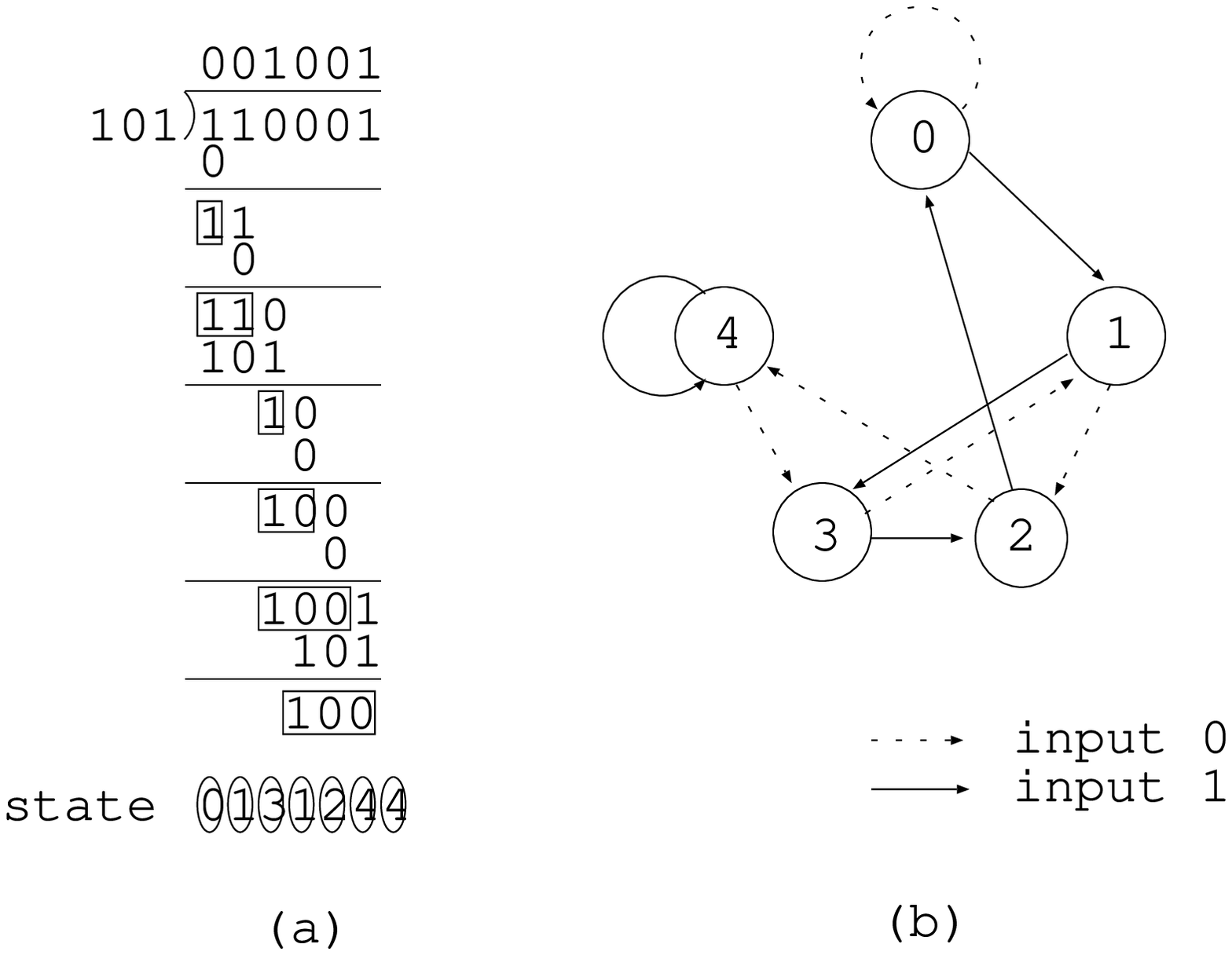}}

\textbf{Fig 2. division procedure for $w_1=110001$ and $p_k=5$}

\vspace{-0.8cm}
  \end{center}
\end{figure}

Let us consider the pfa whose state transition is exactly the
same as $M_0^Q$ of $f(N)$ states excepting that the state
transitions from
$q_{p_k,0,2}$ to $s_l$ for the Fourier transform are replaced by
simple (deterministic) transitions from $q_{p_k,0,2}$ to
$s_N$. We call such a pfa
\textit{emulates} the qfa. Suppose that $M^P$ emulates
$M^Q$. Then the size of $M^P$ is almost the same as that of
$M^Q$, i.e., it is also $\Theta(f(N))$ if the latter is $f(N)$,
since the Fourier transform does not make much difference in the 
number of states. The following lemma is easy:

\textbf{Lemma 4}. Suppose that $M_0^P$ emulates $M_0^Q$. Then
$M_0^P$ accepts strings in $L_0$ with probability one and those
not in $L_0$ with probability $N_0/N$. \hfill \bsquare

Let us set, for example, $N=N_0\sqrt{n}$. Then the error-rate of 
$M_0^Q$ is $(N_0/N)^2=\frac{1}{n}$ and its size is
$O(n^3/\log n)$. To achieve the same error-rate by a pfa, we
have to set $N=N_0n$, which needs $O(n^4/\log n)$ states.

\textbf{Remark}. Suppose that we have once designed a specific
qfa $M_0^Q$ (similarly for $M_0^P$). Then it can work for inputs
of any length or it does not reject the input only by the fact that
its length is not $2n+1$. The above calculation of the acceptance and
rejection rates is only true when our input is restricted to strings
$\subseteq \{0,1\}^n \sharp \{0,1\}^n$.

The following lemmas, Lemma 5 and 6 (see Acknowledgment), are
important for the analysis of error probability of $M_1^Q$, a qfa
which recognizes the second language $L_1(n)$. Here, $\|\psi\|$ means
the norm of a vector $\psi$ and $\|\psi\|_{acc}$ the norm of
$\psi$ after being projected onto accepting space, i.e., the
accepting probability is $\|\psi\|_{acc}^2$. $\left<\psi|\phi\right>$
denotes the inner product between $\psi$ and $\phi$.  

\textbf{Lemma 5}. Let $\psi$ be a quantum state such that applying a
unitary transformation $U$ followed by a measurement to $\psi$  causes
acceptance with probability $1$, i.e., $\|U\psi\|_{acc}^2 = 1$. If
$\psi$ can be decomposed into two orthogonal states $\psi_1$ and
$\psi_2$ s.t. $\psi = \psi_1 + \psi_2$, then $\|U\psi_1\|_{acc}^2 \ge
\|\psi_1\|^4$. 

\textbf{Proof}. Let $H = span\{\phi~\mid~\|U\phi\|_{acc} =
1~and~\|\phi\|=1 \}$, i.e., $H$ is obtained by applying $U^{-1}$ to
the subspace spanned by accepting states only. The accepting probability
of $U\psi_1$ is equal to the squared projection of $U\psi_1$ on the
subspace spanned by accepting states. Since $U$ is unitary and any
unitary transformation preserves the inner product, it turns out that
this projection is the same as the projection of $\psi_1$
onto $H$. Let $H' = span\{\psi\}$. Since $\|U\psi\|_{acc}^2 = 1$, we
have $H' \subseteq H$. Therefore the projection of $\psi_1$ to $H$ is
at least the projection of $\psi_1$ to $H'$, namely at least
$\|\left<\psi_1|\psi\right>\| = \|\psi_1\|^2$. To summarize,
$\|U\psi_1\|_{acc}^2 \ge \|\left<\psi_1|\psi\right>\|^2 =
\|\psi_1\|^4$. \hfill \bsquare   

\textbf{Lemma 6}. Let $\psi$ be a quantum state such that applying a
unitary transformation $U$ followed by a measurement to $\psi$  causes
acceptance with probability at most $\alpha^2$, i.e., $\|U\psi\|_{acc}^2 \le \alpha^2$. If
$\psi$ can be decomposed into two orthogonal states $\psi_1$ and
$\psi_2$ s.t. $\psi = \psi_1 + \psi_2$, then
\begin{equation*}
\|U\psi_1\|_{acc}^2 \le \|\psi_1\|^2\left(\alpha \|\psi_1\| +
\|\psi_2\|\right)^2.
\end{equation*} 

\textbf{Proof}. Let $H$ be the Hilbert space spanned by
$\psi_1$ and $\psi_2$. Then, $\psi_1$ can also be written as  
$$
\psi_1 = \left<\psi|\psi_1\right>\ \psi +
\left<\bar{\psi}|\psi_1\right>\ \bar{\psi},
$$
where $\bar{\psi}$ is a normalized vector in $H$ and
perpendicular to $\psi$. Note that $\|\psi\|$ is also 
$1$. Again $\|\left<\psi|\psi_1\right>\| = \|\psi_1\|^2$ and from the
above equation we obtain that $\left<\psi_1|\psi_1\right> =
\|\left<\psi|\psi_1\right>\|^2 +
\|\left<\bar{\psi}|\psi_1\right>\|^2$, which implies that 
$\|\left<\bar{\psi}|\psi_1\right>\|^2 =
\|\psi_1\|^2(1-\|\psi_1\|^2)$. Note that $\|\psi_1\|^2 + \|\psi_2\|^2 =
1$. Thus, $\|\left<\bar{\psi}|\psi_1\right>\| =
\|\psi_1\|\|\psi_2\|$. Since $ U\psi_1 = \left<\psi|\psi_1\right>\ 
U\psi + \left<\bar{\psi}|\psi_1\right>\ U\bar{\psi}$ and our
observation is a simple projection, it follows by triangular
inequality that   
\begin{eqnarray*}
\|U\psi_1\|_{acc} &\le& \|\left<\psi|\psi_1\right>\| \|U\psi\|_{acc} +
\|\left<\bar{\psi}|\psi_1\right>\| \|U\bar{\psi}\|_{acc}\\
&\le& \|\psi_1\|^2 \alpha + \|\psi_1\|\|\psi_2\| \\
&=& \|\psi_1\| \left(\alpha\|\psi_1\| + \|\psi_2\|\right).
\end{eqnarray*}
This proves the lemma. \hfill \bsquare

Now we shall design a qfa $M_1^Q$ which recognizes the second
language $L_1(n)$. 

\textbf{Lemma 7}. There exists a qfa $M_1^Q$ which accepts
strings in $L_1$ with probability at least $1-(\frai)^2 + (\frai)^4$
and strings not in $L_1$ with at most
$(\frai)^2 + (1-(\frai)^2)(\fraii+\frai)^2$. $M_1^Q$ has
$\Theta((N_1N_2)^2\log N_1 \cdot \log N_2)$ states.  Here $N_0'$
denotes the number $s$ in Lemma 1 but for $x$ and $y$ of length $2n$.

\textbf{Proof}. Again a complete state transition diagram is
shown in Table 2, where accepting states are $s_{N_1,0,p_l,f}$
such that $0\le f \le p_l-1$ and $t_{N_1}$. Rejecting states
are $q_{p_k,e,p_l,f,rej}$ such that $e \neq 0$ or $f \neq 0$, $0
\le e \le p_k-1$, $0 \le f \le p_l-1$, $t_{p_k,0,y}$ such that
$1\le y \le N_2-1$, and $t_z$ such that $1 \le z \le N_1-1$. All
other states are non-halting.

$M_1^Q$ checks whether $\condi$ using $N_1$ primes and also
whether $\condii$ using $N_2$ primes. Note that those two jobs
have to be done at the same time using composite automata while
reading $w_1 \sharp w_2$. Hence $M_1^Q$ first splits into $N_1
\cdot N_2$ submachines, each of which is denoted by $M(k,l)$, $1 
\le k \le N_1$, $1 \le l \le N_2$. As shown in Fig 3, $M(k,l)$
has six stages, from stage 1 thorough stage 6. It might be
convenient to think that each state of $M(k,l)$ be a pair of
state $(q_L,q_R)$ and to think $M(k,l)$ be a composite of $M_L$
and $M_R$. In stages 1 and 2, $M_L$ has similar state
transitions to those of Table 1 for checking $\condin$. $M_R$
has also similar transitions but only for the first part of it,
i.e., to compute $w_1w_2$ mod $p_l$. This portion of transitions 
are given in (2) to (4) of Table 2.
  
\begin{figure}[htbp]
  \begin{center}
   \scalebox{0.55}{\includegraphics{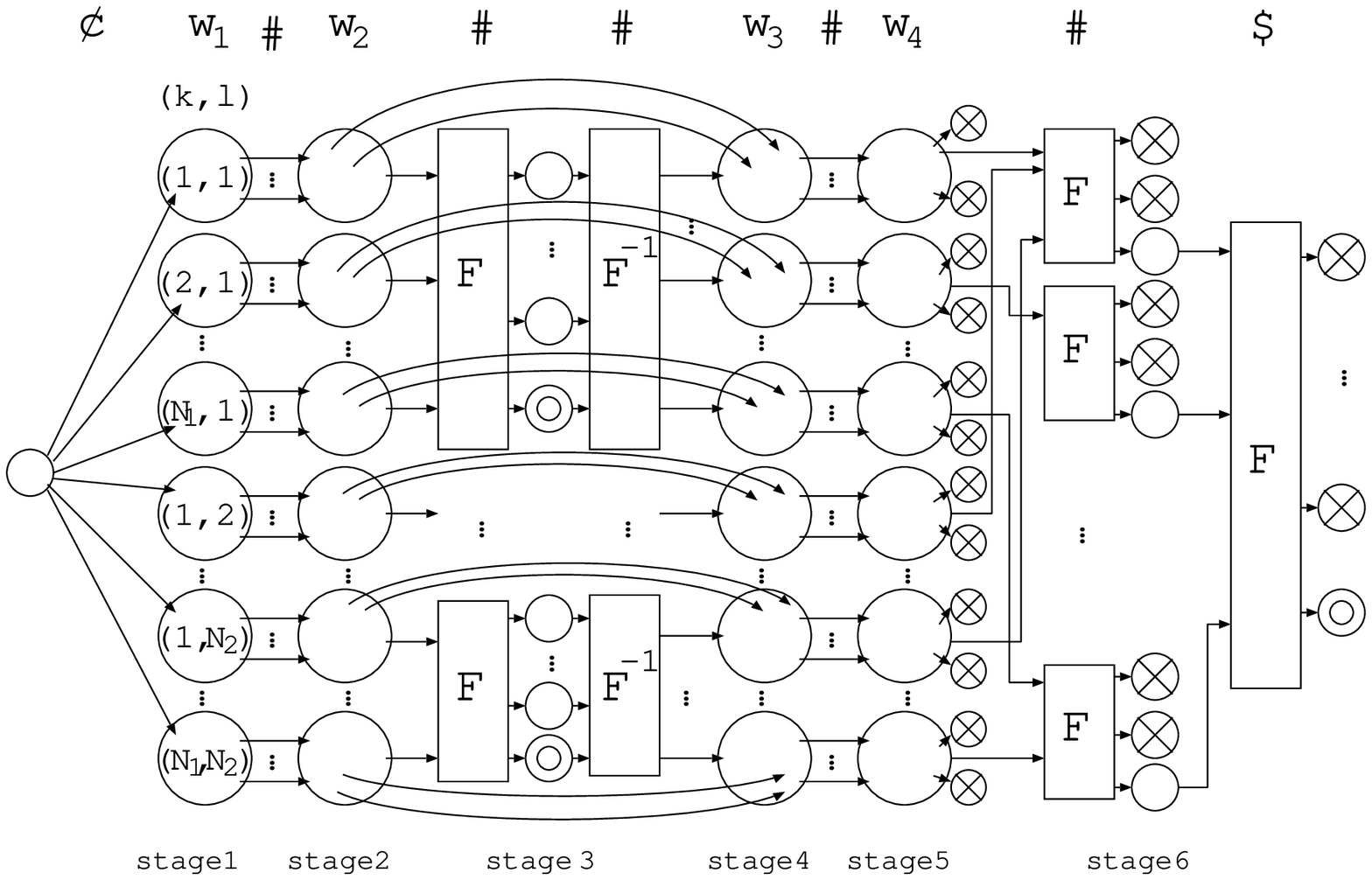}}

   \textbf{Fig 3. qfa $M_1^Q$}

\vspace{-0.8cm}
  \end{center} 
\end{figure}

Now we go to stage 3. Here $M_L$, reading the first $\sharp$,
carries out the Fourier transform exactly as $M_0^Q$ (see
($5-a$) in Table 2). After that $M_L$, reading the second
$\sharp$, execute Inverse Fourier transform from states
$s_{m,0,p_l,f}$ ($1\le m \le N_1$), which is shown
in $(6-a)$ of Table 2. In this stage, $M_R$ does nothing; it
just shifts the state information about $(w_1w_2)$ mod $p_l$
(but only when $\condin$) to stage 4.

Stages 4 and 5 are for the complete reverse operation of stages 2
and 1, respectively. By doing this, the amplitudes for state $q_L$,
which were  
once in turmoil after stage 2, are reorganized and gathered in
specific states, namely $q_{p_k,0,p_l,0,4}$ if
$\condii$. Therefore, what we
do is to gather the amplitude of $q_{p_k,0,p_l,0,4}$ to
$t_{p_k,0,N_2}$ by
Fourier transform reading $\sharp$. Now reading the
rightmost symbol, we do another Fourier transform, which gathers 
the amplitudes of $t_{p_k,0,N_2}$ to $t_{N_1}$.

For the analysis of error probability, Lemma 5 and 6 are
convenient. The basic idea is as follows: When $\condin$, a small
amplitude, $\frac{1}{\sqrt{N_2}}\frac{N_0}{N_1}$ is ``taken'' by each
of the $N_2$ accepting states in stage 3. This is basically the same
as $M_0^Q$ since its probability of observing acceptance is
$\suml\left(\frac{1}{\sqrt{N_2}}\cdot\frai \right)^2 =
\left(\frai\right)^2$. So, the problem is how much of the remaining
amplitudes distributed on the other states in this stage can be
retrieved in the final accepting state $t_{N_1}$ when $\condii$. 

Suppose that we construct a new qfa $M'$ which is
exactly the same as $M_1^Q$ but all the $N_2$ halting states of
$M_1^Q$ in stage 3 are changed to non-halting states. Thus $M'$ only
checks the longer strings, whether $\condii$ or not. It is
clear that $M'$ accepts with probability exactly one when $\condii$ and
with probability at most $(\fraii)^2$ when $\condiin$.  

Note that Lemma 5 and 6 also hold for any sequence of unitary
transformations and measurements since we can delay measurements and
replace them with a single unitary transformation $U$ followed by a
measurement\cite{ANTV99}. Next, consider ($i$)$\psi$, ($ii$)$\psi_1$ and
($iii$)$\psi_2$ in Lemma 5 and 6 as ($i$)the quantum state, ($ii$)the
superposition of non-halting states and ($iii$)the superposition of accepting
states in stage 3 of $M_1^Q$ right after Fourier transform,
respectively. In our case, $\|\psi_1\|^2 = (1-(\frai)^2)$,
$\|\psi_2\|^2 = (\frai)^2$ and $\alpha^2 = (\fraii)^2$. Thus, from
Lemma 5 we can obtain that when $\condii$, $M_1^Q$ accepts with
probability at least  $\left(\frai\right)^2 + \left(1-\left(\frai\right)^2\right)^2.$
Also from Lemma 6, when $\condiin$, $M_1^Q$ accepts with probability at
most
$
\left(\frai\right)^2 +
\left(1-\left(\frai\right)^2\right)\left(\sqrt{\left(1-\left(\frai\right)^2\right)}\fraii+
\frai\right)^2 \le \left(\frai\right)^2 +
\left(1-\left(\frai\right)^2\right)\left(\fraii + \frai\right)^2.  
$

Finally, we count the number of states in $M_1^Q$. This is not
hard since the whole machine is a composition of two machines,
one for using $N_1$ different primes and the other for $N_2$
different primes. Therefore the size of the composite machine is 
$O((N_1N_2)^2 \log N_1 \cdot \log N_2)$. \hfill \bsquare


Suppose that we set $N_1=N_0\sqrt{n}$, and $N_2=dN_0'$. Then
$\frai=\frac{1}{\sqrt{n}}$ and $\fraii + \frai \le \frac{1}{2}$ if we
select a sufficiently large constant $d$, e.g. $d=3$ when $n \ge
36$. Namely, $M_1^Q$ accepts strings in $L_1$ with probability at least
$1-1/n+1/n^2$ and those not in $L_1$ with probability at most
$\frac{1}{4}+\frac{3}{4n}$. The number of states is
$\Theta(\frac{n^5}{\log^2 n})$. The probability distribution for each
state of $M_1^Q$ is illustrated in Fig 4 (for example, above
$1-1/n+1/n^2$ is the sum of $1/n$ and $1-2/n+1/n^2$). 

\begin{figure}[htbp]
  \begin{center}
  \scalebox{0.4}{\includegraphics{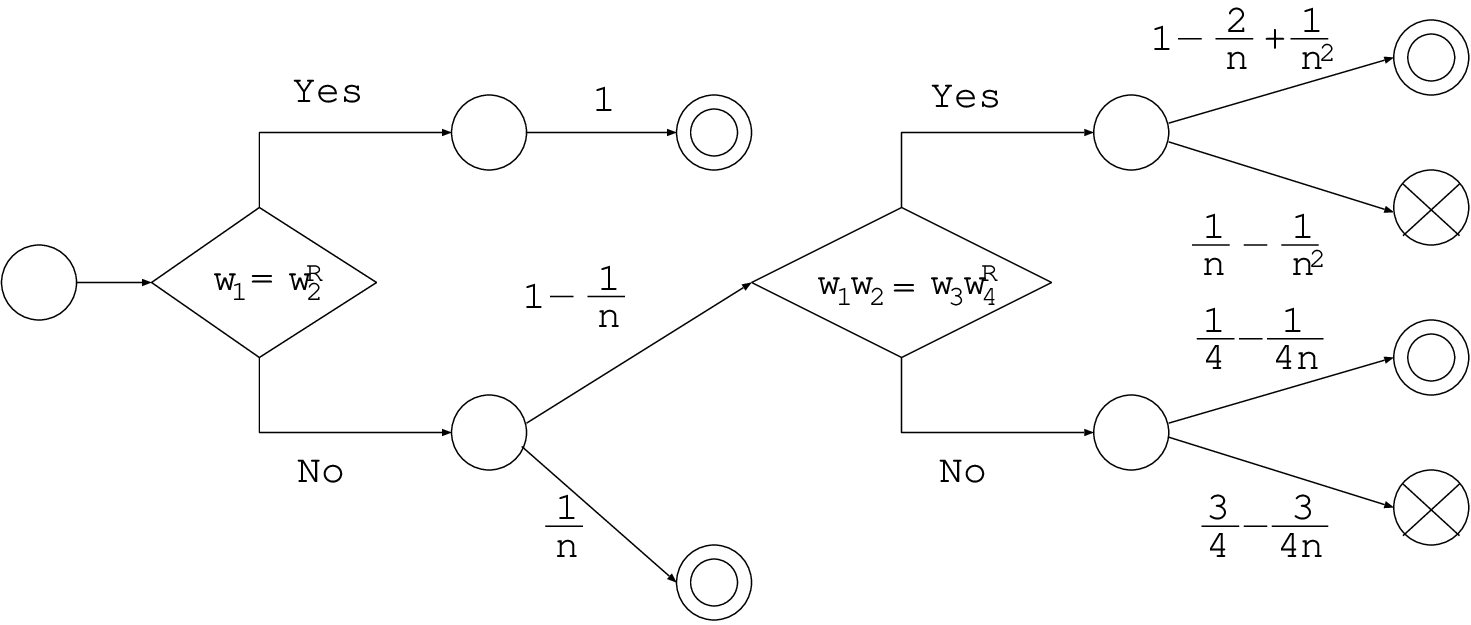}}

  \textbf{Fig 4. probability distribution when
  $N_1=N_0\sqrt{n}$, $N_2=dN_0'$}

\vspace{-0.8cm}
  \end{center}
\end{figure}

Let us consider pfa $M_1^P$ which recognizes $L_1(n)$. The state 
transition of $M_1^P$ is the same as that of $M_1^Q$ except for
Fourier transform and Inverse Fourier transform only $M_1^Q$
performs. If string $x$ satisfies $w_1\neq w_2^R$, then $M_1^P$
accepts $x$ with probability at most $\frac{N_0}{N_1}$ after
reading $w_1\sharp w_2$ instead of with at most $(\frac{N_0}{N_1})^2$
in the case 
of $M_1^Q$. There are subtle differences between $M_1^P$ and
$M_1^Q$. For example, in the case of $M_1^P$, the distributed small
amplitudes after reading $w_3 \sharp w_4$ can be collected completely
(there is some loss due to Inverse Fourier transform in the case of
$M_1^Q$). This causes a slight difference in the accepting probability
of the next lemma (proof is omitted).

\textbf{Lemma 8}. Suppose that $M_1^P$ emulates $M_1^Q$. Then
$M_1^P$ accepts strings in $L_1$ with probability 1 and
those not in $L_1$ with probability at most
$\frac{N_0}{N_1}+\left(1-\frai\right)\cdot
\frac{N_0'}{N_2}$. The number of states is 
approximately the same as the one of $M_1^Q$, i.e.,
$\Theta((N_1N_2)^2\log N_1 \log N_2)$. \hfill \bsquare

If we set $N_1=N_0n$ and $N_2=dN_0'$, then strings such that
$w_1 \neq w_2^R$ are accepted with probability at most
$\frac{1}{n}$ after reading $w_1\sharp w_2$. Thus this
probability is the same as qfa $M_1^Q$ such that $N_1=N_0\sqrt{n}$
and $N_2=dN_0'$, but the lemma says that we need
$\Omega\left(\frac{n^6}{\log^2n}\right)$ states. See Fig. 5 for a
probability distribution. 

\begin{figure}[htbp]
\begin{center}
  \scalebox{0.4}{\includegraphics{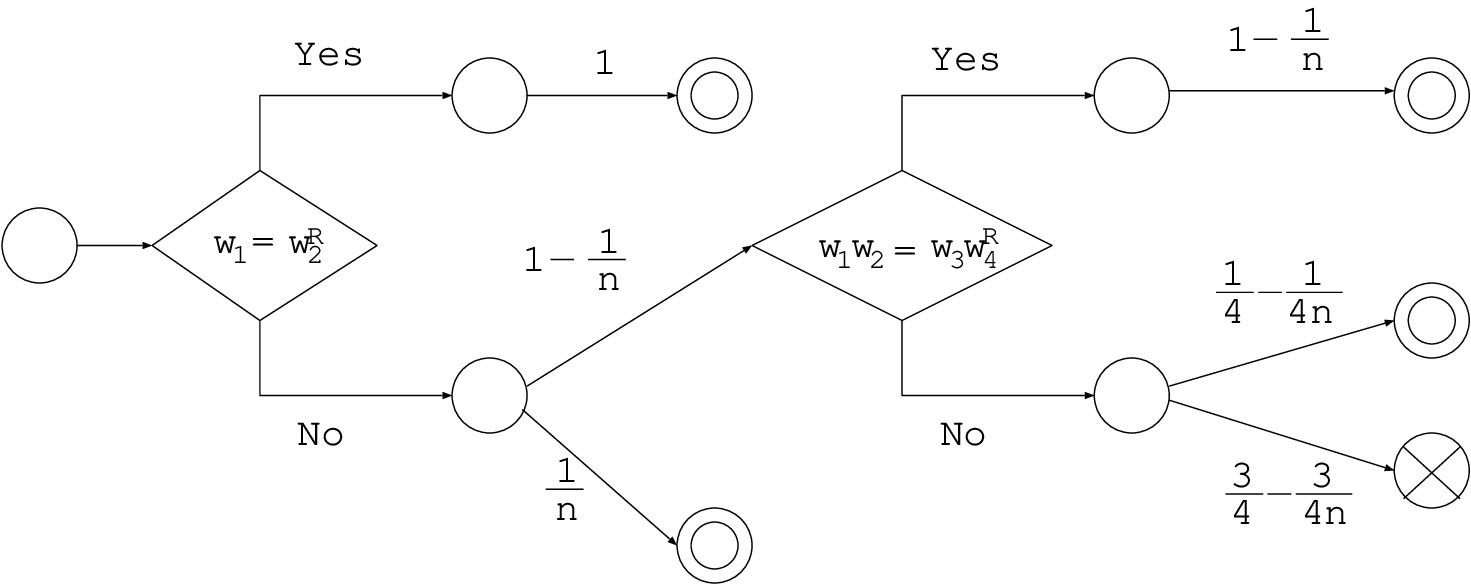}}

  \textbf{Fig 5. probability distribution when
  $N_1=N_0n$, $N_2=dN_0'$}
\vspace{-0.8cm}
\end{center}
\end{figure}

Now we are ready to give our main theorem:

\textbf{Theorem 1}. For any integer $c$, there is a qfa
$M^Q$ such that $M^Q$ recognizes $L(n,n^c)$ and the number of
states in $M^Q$ is $O\left(\frac{n^{c+4}}{\log^2n}\right)$.

\textbf{Proof}. The construction of $M^Q$ is easy: We just add a 
new deterministic transition from the last accepting state in
stage 6 of $M_1^Q$ to its initial state by reading $\sharp$, by which
we can
manage iteration. Also, we need some small changes to handle the
very end of the string: Formally speaking, transition (11) in
Table 2 is modified into
$$V_\sharp|t_{p_k,0,N_2}\rangle=\frac{1}{\sqrt{N_1}}
\sum_{z=1}^{N_1}\exp\left(\frac{2\pi i}{N_2}kz\right)
|t_z\rangle,$$
$t_{N_1}$ is now not an accepting state but a non-halting state
and two new transitions
$$
\begin{array}{lc}
(10-c) &
V_\$|q_{p_k,e,p_l,f,4}\rangle=|q_{p_k,e,p_l,f,rej}\rangle\\
(12) & \quad V_\sharp|t_{N_1}\rangle=|q_1\rangle
\end{array}
$$
are added.

We set $N_1=2N_0n^{c/2}$ and $N_2=dN_0'$. Then
$N_0/N_1=\frac{1}{2n^{c/2}}$ and $\fraii + \frai < \frac{1}{2}$ if we
select a sufficiently large constant as $d$. Suppose that $M^Q$ has not
halted yet and is now reading the $i$-th block $w_{i1}\sharp
w_{i2} \sharp \sharp w_{i3} \sharp w_{i4}$. Then, we can
conclude the following by Lemma 7: $(i)$ If $w_{i1}=w_{i2}^R$, then
$M^Q$ accepts the input with probability one. ($ii$) If
$(w_{i1}w_{i2})=(w_{i3}w_{i4})^R$, then $(ii-a)$ $M^Q$ also
accepts the input with probability at most $1/4n^c$ and $(ii-b)$ rejects the
input with at most $\frac{1}{4n^c}-\frac{1}{16n^{2c}}$ and $(ii-c)$ goes 
back to the initial state with at least $1-\frac{2}{4n^c}+
\frac{1}{16n^{2c}}$. $(iii)$ If
$(w_{i1}w_{i2})\neq(w_{i3}w_{i4})^R$, then $(iii-a)$ $M^Q$ accepts
the input with at most $\frac{1}{4n^c}$, $(iii-b)$ rejects it with
at least $\frac{3}{4}-\frac{3}{16n^c}$ and $(iii-c)$
goes back to the initial state with at most 
$\frac{1}{4}-\frac{1}{16n^c}$. The number of state is
$O(n^{c+4}/\log^2n)$. 

Recall that the number of iteration is $n^c$. Now suppose that
the input $x$ is in
$L(n,n^c)$. Then, the probability that $x$ is rejected is equal
to the
probability that $(ii-b)$ happens before ($i$) happens. The
probability that $(ii-b)$ happens is at most $\frac{1}{4n^c}$ per 
iteration, and so the probability that ($ii-b$) happens in some
iteration is at most $n^c \cdot
\frac{1}{4n^c}=\frac{1}{4}$. Therefore the probability that $x$
is finally accepted is well larger than 1/2. Suppose conversely
that $x$ is not in $L(n,n^c)$. Then the probability that
$(ii-a)$ happens in some iteration is the same as 
above and is at most $\frac{1}{4}$. If $M^Q$ does not meet a
block such that $(w_{i1}w_{i2})\neq(w_{i3}w_{i4})^R$ until the
end, then the accepting probability is at most this $1/4$. If
$M^Q$ does meet such a block in some iteration, it rejects $x$ with
probability at least $(1-\frac{1}{4})(\frac{3}{4}-
\frac{3}{16n^c})$ which is again well above 1/2. Thus $M^Q$ recognizes
$L(n,n^c)$. \hfill \bsquare 

\textbf{Theorem 2}. Suppose that $M^P$ which emulates $M^Q$
recognizes $L_2(n,n^c)$. Then the number of states of $M^P$ is
$\Omega(n^{2c+4}/\log^2n)$.

\textbf{Proof}. $M^P$ is constructed by applying the same 
modification (as given in the above proof) to $M_1^P$. Then it turns
out that we must set $N_0'/N_2 \le 1/d$, where $d$ is
a sufficiently large constant, to reject the strings such that 
$(w_{i1}w_{i2}) \neq (w_{i3}w_{i4})^R$ since $M_1^P$ accepts such bad
strings with probability at least $\frac{N_0}{N_1} +  
(1-\frac{N_0}{N_1})\cdot \frac{N_0'}{N_2}$ by Lemma 8. So we have to
set $N_2 = dN_0'$ and suppose that we set
$N_1=\frac{1}{a}N_0n^c$. Then, as shown below, $M^P$ does not   
recognize $L_2(n,n^c)$ if $a$ is large. That means we have to set  
$N_1=\frac{1}{a}N_0n^c$ for a sufficiently small $a>0$, which implies,
from Lemma 8, that we need $\Omega(n^{2c+4}/\log^2n)$ states. 

Now suppose
that the input $x$ includes a long repetition of blocks such
that $(w_{i1}w_{i2})=(w_{i3}w_{i4})^R$. Then $x$ is accepted in
each iteration with probability $a/n^c$. Therefore the
probability that this happens in the first $k$ iterations is
$$\sum_{i=1}^k\left(1-\frac{a}{n^c}\right)^{i-1}\cdot \frac{a}{n^c}
=1-\left(1-\frac{a}{n^c}\right)^k.$$
Since the number of repetitions $(=k)$ can be as large as $n^c$, 
$$\lim_{n\to\infty}\left(1-\frac{a}{n^c}\right)^{n^c}=\frac{1}{e^a}.$$
Thus if we select a sufficiently large constant $a$, then the
probability of acceptance can be arbitrarily close to one. Such an $M^P$
does not recognize $L(n,n^c)$ obviously, which proves the
theorem. \hfill \bsquare

\vspace{-5mm}
\section{Concluding Remarks}
\vspace{-2mm}

The question in this paper is whether or not we can exploit the
difference in probability calculation between quantum and
probabilistic computations. We have shown that the answer is yes 
using quantum finite automata. However, what remains apparently
is whether or not we can exploit this property for other types
of models and/or for other types of problems which are
preferably less artificial. Also it should be an important
future research to obtain a general lower bound for the number
of states which is needed to recognize $L_2(n,n^c)$ by pfa's.

\textbf{Acknowledgment}. We are grateful to Mario Szegedy for his
many valuable comments to this research. We also thank the anonymous
reviewer for Lemma 5. In the earlier version of this paper, the proof
of Lemma 7 was very lengthy. Using this lemma and Lemma 6 which is
developed under the same idea as Lemma 5, the proof of Lemma 7 was
greatly simplified.

\begin{small} 
$$
\begin{array}{ll}
(1) & V_{\cent}|q_0\rangle = \frac{1}{\sqrt{N_1N_2}}
\sumk\suml|q_{p_k,0,p_l,0,1}\rangle, \\
(2-a) & V_0\qi = |q_{p_k,2e,p_l,2f,1}\rangle \quad (\ei,\quad \ffi),\\
(2-b) & V_0\qi = |q_{p_k,2e,p_l,2f-p_l,1}\rangle \quad (\ei,\quad
\ffii), \\
(2-c) & V_0\qi = |q_{p_k,2e-p_k,p_l,2f,1}\rangle \quad (\eii,\quad
\ffi),\\
(2-d) & V_0\qi = |q_{p_k,2e-p_k,p_l,2f-p_l,1}\rangle \quad (\eii,\quad
\ffii),\\
(2-e) & V_1\qi = |q_{p_k,2e+1,p_l,2f+1,1}\rangle \quad (\eiii,\quad
\ffiii),\\
(2-f) & V_1\qi = |q_{p_k,2e+1,p_l,2f+1-p_l,1}\rangle \quad (\eiii,\quad
\ffiiii),\\
(2-g) & V_1\qi = |q_{p_k,2e+1-p_k,p_l,2f+1,1}\rangle \quad (\eiiii,\quad
\ffiii),\\
(2-h) & V_1\qi = |q_{p_k,2e+1-p_k,p_l,2f+1-p_l,1}\rangle \\
& \hspace{3cm}(\eiiii,\quad \ffiiii),\\
(3) & V_{\sharp}\qi = \qii, \\
(4-a) & V_0\qii = |q_{p_k,\frac{e}{2},p_l,2f,2}\rangle \quad
(\eeven,\quad \ffi),\\ 
(4-b) & V_0\qii = |q_{p_k,\frac{e}{2},p_l,2f-p_l,2}\rangle \quad
(\eeven,\quad \ffii), \\
(4-c) & V_0\qii = |q_{p_k,\frac{e+p_k}{2},p_l,2f,2}\rangle \quad
(\eodd,\quad \ffi),\\
(4-d) & V_0\qii = |q_{p_k,\frac{e+p_k}{2},p_l,2f-p_l,2}\rangle \quad
(\eodd,\quad \ffii),\\
(4-e) & V_1\qii = |q_{p_k,\frac{e-1+p_k}{2},p_l,2f+1,2}\rangle \quad
(\eeven,\quad \ffiii),\\
(4-f) & V_1\qii = |q_{p_k,\frac{e-1+p_k}{2},p_l,2f+1-p_l,2}\rangle
\quad (\eeven,\quad \ffiiii),\\
(4-g) & V_1\qii = |q_{p_k,\frac{e-1}{2},p_l,2f+1,2}\rangle \quad
(\eodd,\quad \ffiii),\\
(4-h) & V_1\qii = |q_{p_k,\frac{e-1}{2},p_l,2f+1-p_l,2}\rangle \quad
(\eodd,\quad \ffiiii),\\
(5-a) & V_\sharp|q_{p_k,0,p_l,f,2}\rangle =\frac{1}{\sqrt{N_1}}
\sum_{m=1}^{N_1}\exp\left(\frac{2\pi
i}{N_1}km\right)|s_{m,0,p_l,f}\rangle, \\   
(5-b) & V_\sharp\qii=|q_{p_k,e,p_l,f}\rangle \quad (1 \le e < p_k), \\
(6-a) & V_\sharp|s_{m,0,p_l,f}\rangle =\frac{1}{\sqrt{N_1}}
\sum_{r=1}^{N_1}\exp\left(-\frac{2\pi
i}{N_1}mr\right)|q_{p_r,0,p_l,f,3}\rangle \ (1 \le m \le N_1),\\
(6-b) & V_\sharp |q_{p_k,e,p_l,f}\rangle =\qiii \quad (1 \le e <
p_k),\\
(7-a) & V_0\qiii = |q_{p_k,2e,p_l,\frac{f}{2},3}\rangle \quad
(\ei,\quad \feven),\\ 
(7-b) & V_0\qiii = |q_{p_k,2e,p_l,\frac{f+p_l}{2},3}\rangle \quad
(\ei,\quad \fodd), \\
(7-c) & V_0\qiii = |q_{p_k,2e-p_k,p_l,\frac{f}{2},3}\rangle \quad
(\eii,\quad \feven),\\
(7-d) & V_0\qiii = |q_{p_k,2e-p_k,p_l,\frac{f+p_l}{2},3}\rangle \quad
(\eii,\quad \fodd),\\
(7-e) & V_1\qiii = |q_{p_k,2e+1,p_l,\frac{f-1+p_l}{2},3}\rangle \quad
(\eiii,\quad \feven),\\
(7-f) & V_1\qiii = |q_{p_k,2e+1,p_l,\frac{f-1}{2},3}\rangle \quad
(\eiii,\quad \fodd),\\
(7-g) & V_1\qiii = |q_{p_k,2e+1-p_k,p_l,\frac{f-1+p_l}{2},3}\rangle
\quad (\eiiii,\quad \feven),\\
(7-h) & V_1\qiii = |q_{p_k,2e+1-p_k,p_l,\frac{f-1}{2},3}\rangle \quad
(\eiiii,\quad \fodd),\\
(8) & V_{\sharp}\qiii = \qiiii,\\
(9-a) & V_0\qiiii = |q_{p_k,\frac{e}{2},p_l,\frac{f}{2},4}\rangle \quad
(\eeven,\quad \feven),\\ 
(9-b) & V_0\qiiii =|q_{p_k,\frac{e}{2},p_l,\frac{f+p_l}{2},4}
\rangle \quad (\eeven,\quad \fodd), \\
(9-c) & V_0\qiiii = |q_{p_k,\frac{e+p_k}{2},p_l,\frac{f}{2},4}\rangle
\quad (\eodd,\quad \feven),\\
(9-d) & V_0\qiiii =|q_{p_k,\frac{e+p_k}{2},p_l,p_l,\frac{f+p_l}{2},4}
\rangle \quad (\eodd,\quad \fodd),\\
(9-e) & V_1\qiiii =|q_{p_k,\frac{e-1+p_k}{2},p_l,\frac{f-1+p_l}{2},4}
\rangle \quad (\eeven,\quad \feven),\\
(9-f) & V_1\qiiii = |q_{p_k,\frac{e-1+p_k}{2},p_l,\frac{f-1}{2},4}
\rangle \quad (\eeven,\quad \fodd),\\
(9-g) & V_1\qiiii =|q_{p_k,\frac{e-1}{2},p_l,\frac{f-1+p_l}{2},4}
\rangle \quad (\eodd,\quad \feven),\\
(9-h) & V_1\qiiii = |q_{p_k,\frac{e-1}{2},p_l,\frac{f-1}{2},4}\rangle
\quad (\eodd,\quad \fodd),\\
(10-a) & V_\sharp|q_{p_k,0,p_l,0,4}\rangle =\frac{1}{\sqrt{N_2}}
\sum_{y=1}^{N_2}\exp\left(\frac{2\pi i}{N_2}ly\right)
|t_{p_k,0,y}\rangle, \\
(10-b) & V_\sharp\qiiii=|q_{p_k,e,p_l,f,rej}\rangle \quad (1 \le f <
p_l), \\
(11) & V_\$|t_{p_k,0,N_2}\rangle=\frac{1}{\sqrt{N_1}}
\sum_{z=1}^{N_1}\exp\left(\frac{2\pi i}{N_2}kz\right) |t_z\rangle, \\
\end{array}
$$
\end{small}
  \begin{center}
\textbf{Table 2.} state transition diagram of $M_1^Q$
  \end{center}

\end{document}